\documentclass[a4paper]{ESASPCS13Style}
\usepackage{epsfig}

\begin{document}

\title{Attacking the X-ray emission properties of young stars with the Sword of Orion}

\author{K.\,R. Briggs\inst{1} \and M. G\"udel\inst{1} \and
  M. Audard\inst{2} \and K. Smith\inst{3}} 
\institute{Paul Scherrer Institut, CH-5232 Villigen PSI, Switzerland
  \and Columbia Astrophysics Laboratory, 550 West 120th Street, New York, NY~10027, USA \and Max-Planck-Institut f\"ur Radioastronomie, Auf dem H\"ugel 69,
  D-53121, Bonn, Germany } 

\maketitle 

\begin{abstract}

We present a survey of X-ray emission from young stars in the Sword of
Orion star-formation region using {\it XMM-Newton}'s EPIC detectors. We
find over 850 X-ray sources, of which more than 700 have near-infrared
counterparts consistent with being young stars. The survey enables
statistical investigation of the dependence of X-ray emission
properties of young stars on fundamental stellar parameters (mass,
rotation, age) and environmental features (circumstellar disk, active
accretion, circumstellar absorption), and study of structure size in
individual coronae through analysis of large flares.

\keywords{Stars: activity -- Stars: coronae -- Stars: flare -- Stars: pre-main sequence -- X-rays: stars\ }
\end{abstract}

\section{Introduction}

A young star grows within a complex environment, accreting from a circumstellar
disk and driving jets out through a surrounding envelope of dust and
molecular gas. Fast rotation and a fully-convective structure also distance
the emerging star itself from the familiar Sun. The impact of these
circumstances upon a stellar magnetic dynamo, the structure of
magnetic field on the star and in its immediate environment, and
resultantly the location, temperature, and amount of X-ray-emitting hot
plasma within the system is yet to be well established. As the X-ray
emission is in turn expected to be the key agent for ionization of
circumstellar material, understanding the interaction of these
processes is important in understanding the early stages of stellar
evolution.

Studies of the X-ray emission from stars in star-forming regions
(SFRs) have produced conflicting results as to whether young stars do
(e.g. \cite{sn01}) or do not (e.g. \cite{feig03}; \cite{flac03}) 
demonstrate the same dependence of X-ray luminosity on
rotation as main-sequence stars, and as to whether accretion or the
presence of a circumstellar disk does (e.g. \cite{sn01};
\cite{flac03}) or does not (e.g. \cite{pz02}; \cite{feig03}) suppress 
the observed X-ray luminosity. The conflict has been variously
attributed to: real physical differences between different SFRs;
biases in sample selection; and incorrect assumptions (e.g. source
spectrum, absorbing column density) in calculating X-ray luminosities.

Our survey of the Sword of Orion examines a region of star formation
intermediate in star density and mass distribution between the
rich, Orion Nebula Cluster (ONC: \cite{feig03};
\cite{flac03}), centrally concentrated on massive O-type stars whose
strong UV radiation may affect evolution of circumstellar material of
lower-mass stars, and the sparser Taurus molecular cloud
(\cite{sn01}), lacking in high-mass stars. 

In this paper, we show preliminary results from the survey, including the
X-ray and near-infrared (NIR) source content of the field, and
examples of spectral and temporal analyses of individual sources.

\section{Observations and analysis}
\label{sec:obs}

\begin{figure}[h!]
\centering{
\includegraphics[height=0.82\textheight, origin=c, angle=0]{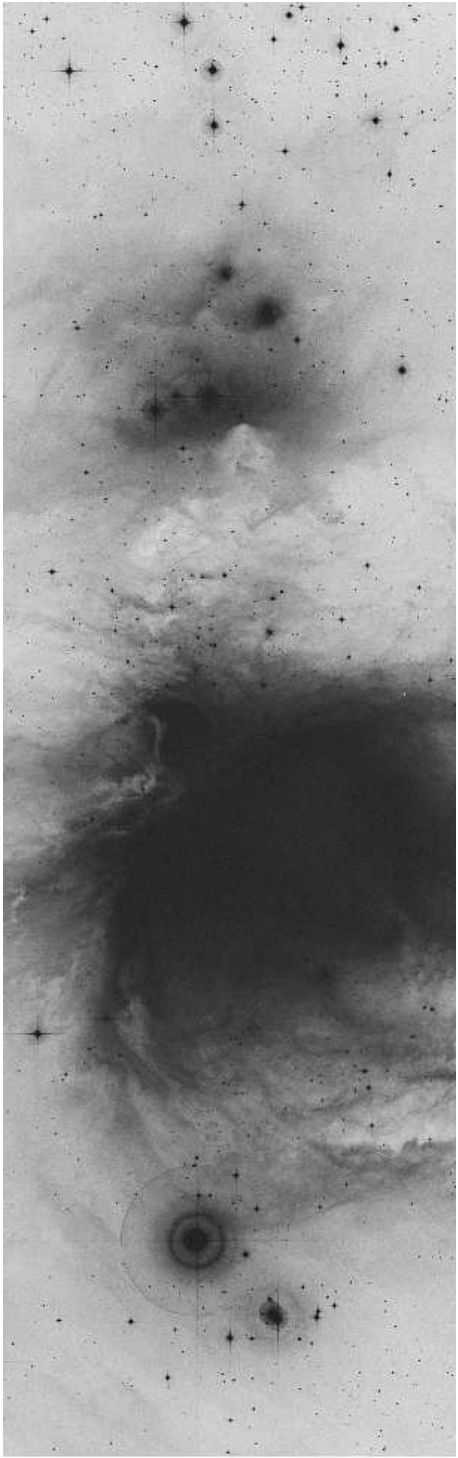}
}
\caption{{\it Optical (red) DSS image of the survey area.}}
\label{fig_opt_image}
\end{figure}

\begin{figure}[h!]
\centering{
\includegraphics[height=0.82\textheight, origin=c, angle=0]{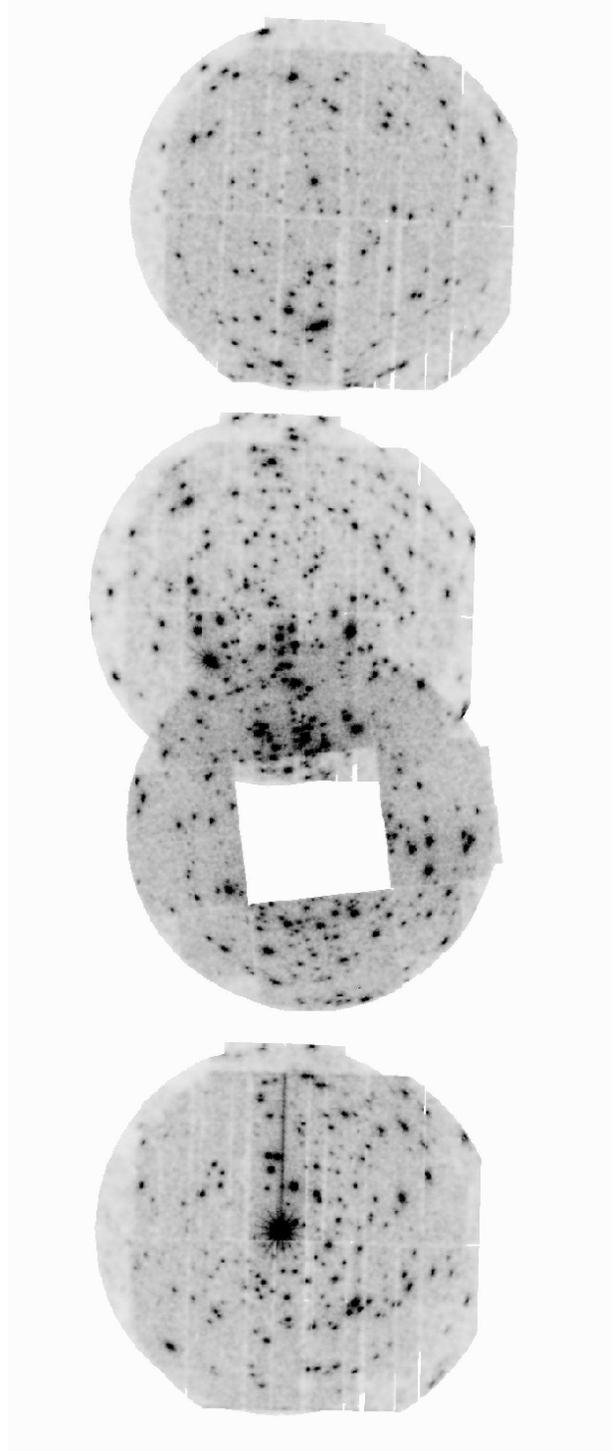}
}
\caption{{Mosaicked EPIC X-ray image (0.3--4.5 keV) of the survey area
    (to same scale as Fig.~\protect\ref{fig_opt_image}).}}
\label{fig_epic_image}
\end{figure}

The survey consists of four {\it XMM-Newton} fields covering a region
$\sim 2$\degr$ \times 0.5$\degr. An optical image (Fig.~\ref{fig_opt_image})
of the survey area shows the extensive nebulosity, including the Orion
Nebula itself, associated with the star-formation activity in this
region. In three fields all three EPIC instruments (PN and both MOS
cameras) were exposed in Full-Frame mode for $\approx 20$ ks; in the
field centred on the ONC 
(\object{$\theta^1$~Ori~C}) itself only the MOS cameras were exposed in Partial
Window mode for 36 ks. Each event-list was filtered to exclude events
with bad flags and patterns, and times of high background. For each
observation, an image in the 0.3--4.5 keV band was created for each
operating instrument, and these were mosaicked to make an EPIC
image. Source detection was performed using the Science Analysis System
(SAS\footnote{http://xmm.vilspa.esa.es})
tasks {\sc ewavelet} and {\sc emldetect}. 
Figure~\ref{fig_epic_image} shows a mosaic of the EPIC survey images.
The optically-brightest star in the survey, \object{$\iota$~Ori}, is
the bright X-ray source in the southernmost survey field. 

\section{Results}
\label{sec:res}

\subsection{Characterization of X-ray sources}

\begin{figure}[ht]
\centering{
\includegraphics[height=0.85\columnwidth, origin=c, angle=270]{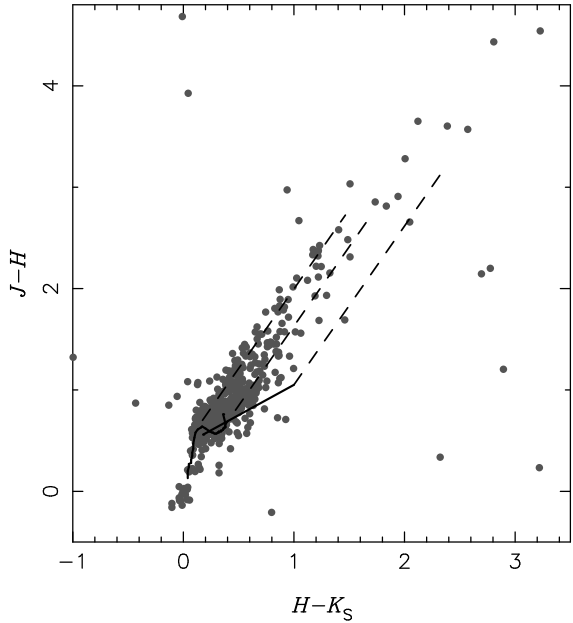}
}
\caption{2MASS $JHK$ two-colour diagram of objects coincident with
  EPIC-detected X-ray sources in the survey. The bold curve marks the
expected locus of main-sequence stars (\protect\cite{bar98}); the bold
straight line marks the locus of young stars with circumstellar disks
(\protect\cite{meyer97}); the dashed lines indicate expected reddening
vectors, extending to $A_{\rm V} = 20$.}
\label{fig_2mass}
\end{figure}

We have detected over 850 X-ray sources in the survey. We performed
cross-correlation of the X-ray source list with 2MASS sources in the
survey area. More than 700 have 2MASS counterparts within 6 arcsec,
the vast majority having the photometric properties of young stars
(Fig.~\ref{fig_2mass}). A small fraction of the X-ray-detected stars
show the $K$-band excess expected of circumstellar disks; a 
few show extreme $K$-band excess indicative of further circumstellar
material and an early stage (Class I) of evolution.

Much work has been done to characterize the stars in this region by
spectral type, mass, age, reddening, rotation period, circumstellar
environment (e.g. \cite{hil97}; \cite{reb00}; \cite{reb01};
\cite{chs01}). By studying the dependence of X-ray
luminosity, $L_{\rm X}$, (particularly normalized to a star's
bolometric luminosity, $L_{\rm bol}$) on these parameters we aim to
understand the mechanisms driving the X-ray and magnetic activity of
these young stars. 

\subsection{The rotation--activity relation}

Analysis of the northernmost field in the survey found no 
clear dependence of $L_{\rm X}$ or $L_{\rm X}/L_{\rm bol}$ on rotation
period (\cite{krb04}), in contrast to that of MS solar-like stars. 
This suggests a different dynamo mechanism is at work. However, the
Rossby numbers (ratio of rotation and convective turnover periods) of
these fully-convective stars are consistent with those of
saturated-emission solar-like stars, whose high levels of activity -- $L_{\rm
  X}/L_{\rm bol} \sim 10^{-3}$ -- also show no rotation dependence. 

\subsection{The influence of accretion}

Analysis of this subsample also found that low-mass stars ($M < 0.5$
M$_{\odot}$) showing an excess of $U$-band continuum emission
(\cite{reb00}), expected 
from material that is heated as it accretes onto a star, had lower
median $L_{\rm X}/L_{\rm bol}$, by a factor $\approx 2$, than stars showing no
such significant excess (\cite{krb04}). This implies that
accretion plays a role in suppressing the observed X-ray emission.

\begin{figure}[t]
\centering{
\includegraphics[height=0.75\columnwidth, origin=c, angle=270]{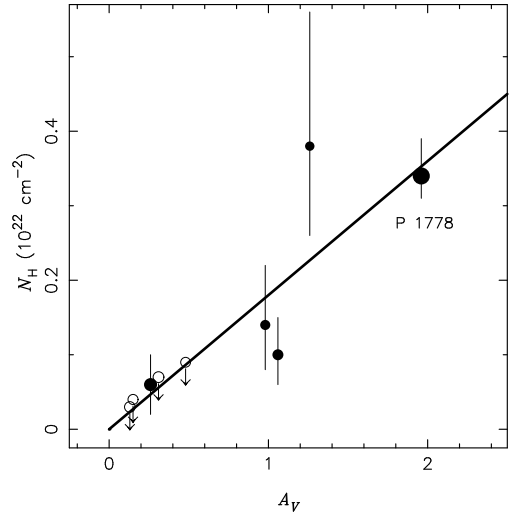}
}
\caption{Fitted absorbing column density, $N_{\rm H}$, vs. visual
  extinction, $A_{\rm V}$. The line shows the standard relation
  (Paresce 1984). Open circles show upper limits.}
\label{fig_nh}
\end{figure}

\cite*{stas04} have proposed that columns of accreting material obscure 
underlying coronal emission, i.e. the absorbing column to X-rays,
$N_{\rm H}$ is higher than would be anticipated from the observed
visual extinction, $A_{\rm  V}$. We aim to test this though
spectroscopic analysis of strongly- and weakly/non-accreting samples. In a
small sample studied so far, we have found the $N_{\rm H}$ derived
from spectral fitting is no more than that expected from optical
absorption (Fig.~\ref{fig_nh}).

\subsection{Spectral analysis of a young solar analog}

Through spectroscopic analysis, we also aim to assess the temperature
of the X-ray-emitting plasma of bright sources and investigate its
dependence on rotation period and evolutionary stage. Solar-like MS
stars show increasing coronal temperature with decreasing rotation
period (or Rossby number). \cite*{tsuj02} found that 
X-ray-bright young stars in earlier evolutionary stages 
(Class I--II) had higher plasma
temperatures than those in later stages (Class III). Few 
Class II objects have been studied with high-resolution X-ray
spectroscopy: while \object{SU~Aur} (\cite{kester04}) and
\object{RY~Tau} (\cite{audard04}) also show hot temperatures,
\object{TW~Hya} has remarkably cool plasma (\cite{kast02}; \cite{ss04}). 

\begin{figure}[t]
\centering{
\includegraphics[height=0.8\columnwidth, origin=c, angle=270]{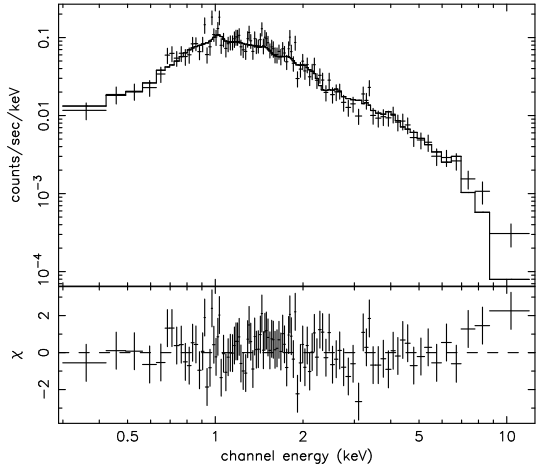}
}
\caption{{\it PN spectrum of young solar analog Parenago~1778.}}
\label{fig_spec_p1778}
\end{figure}

Figure~\ref{fig_spec_p1778} shows a PN spectrum of \object{Parenago
  1778}, an approximately 0.75~Myr-old solar analog (according to the
models of \cite{siess00}). 
The star's 2MASS photometry shows small excesses in the $H$ and $K$
bands, suggestive of a small circumstellar disk, but there is a strong
$U$-band excess (\cite{reb00}) indicative of a high accretion rate
($\dot{M} \sim 3~\times~10^{-8}$~M$_{\odot}$\,yr$^{-1}$). 
The X-ray luminosity is very high, $L_{\rm X} \approx
4~\times~10^{31}$~erg\,s$^{-1}$, with $\log(L_{\rm X}/L_{\rm bol})
\approx -2.4$. A two-temperature fit to the PN spectrum, using the
elemental abundances of a very active near-MS star (\object{AB~Dor};
\cite{sanz03}) requires the hot component to be at $\approx 50$~MK, a
temperature observed on MS stars only transiently in very large
flares, yet the PN lightcurve of \object{Parenago~1778} does not
indicate that such a flare occured during this observation.

\begin{figure}[t]
\centering{
\includegraphics[height=0.77\columnwidth, origin=c, angle=270]{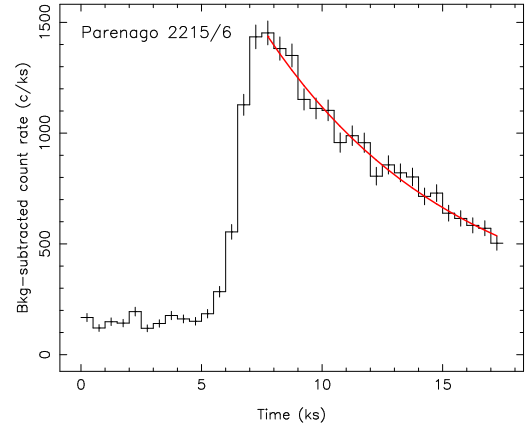}
}
\vspace{-0.025\textheight}
\caption{{\it PN lightcurve of the system Parenago~2215/2216.}}
\label{fig_lc_p2215}
\end{figure}

A number of such flares are observed in the survey
(e.g. Fig.~\ref{fig_lc_p2215}). Hydrodynamic modelling of the
temperature and emission measure of the flare plasma during the decay
phase allows estimation of the size of magnetic loop involved
(\cite{rm98}). Loop sizes larger than the stellar radius may indicate
the flare occurred in magnetic structure bridging the star and a
circumstellar disk, or within the disk itself, rather than in a
coronal loop on the star. 

\section{Summary}

An {\it XMM-Newton} survey of the Sword of Orion star-forming region
has yielded over 850 X-ray sources, of which more than 700 have near-infrared
counterparts consistent with being young stars. Analysis of a
subsample of the survey has indicated that these young stars do not
show the activity--rotation relation of solar-like MS stars, and that
accretion suppresses the observed X-ray emission of young
stars. Absorption of X-rays does not appear to be larger than expected
from observed optical extinction. Extremely hot plasma temperatures of
$\approx 50$~MK are found on at least one young solar
analog. Statistical investigations, using the whole sample, of the
dependence of X-ray emission levels and plasma temperatures on
fundamental stellar parameters and environmental features are being
pursued, and detailed analysis of large flares aims to identify the
emission region in these systems.

\begin{acknowledgements}

Financial support from the Swiss NSF (grant 20-66875.01) is acknowledged.
This research is based on observations obtained with {\it XMM-Newton},
an ESA mission funded by ESA Member States and the USA (NASA), and
makes use of data products from the Two Micron All 
Sky Survey, which is a joint project of the University of
Massachusetts and the Infrared Processing and Analysis
Center/California Institute of Technology, funded by the National
Aeronautics and Space Administration and the National Science
Foundation. The Digitized Sky Surveys were produced at the Space
Telescope Science Institute under U.S. Government grant NAG
W-2166; the image presented herein was based on photographic data
obtained by the UK Schmidt Telescope, operated by the Anglo-Australian
Observatory.

\end{acknowledgements}

\end{document}